\documentclass[journal]{IEEEtran}

\ifCLASSINFOpdf
  \usepackage[pdftex]{graphicx}
\else
\fi

\usepackage{amsmath,amsfonts,amssymb,bm,color,cite}
\usepackage{times}
\usepackage{graphicx}
\usepackage{url}
\usepackage{booktabs}
\usepackage{caption,subcaption}

\usepackage{array}

\usepackage{multicol}
\usepackage{multirow}

\usepackage{url}

\begin{document}
%
\title{Explainable, Physics Aware, Trustworthy AI Paradigm Shift for Synthetic Aperture Radar}
%
%
%
\author{Mihai~Datcu,~\IEEEmembership{Fellow,~IEEE}~,
        Zhongling~Huang$^\dag$,
        Andrei~Anghel,
        Juanping~Zhao,
        Remus~Cacoveanu
\thanks{\textit{$^\dag$ Corresponding author (huangzhongling@nwpu.edu.cn)}}
}

%
%

\markboth{Accepted Version}%
{Shell \MakeLowercase{\textit{et al.}}: Bare Demo of IEEEtran.cls for IEEE Journals}
%



\maketitle

\begin{abstract}
The recognition or understanding of the scenes observed with a SAR system requires a broader range of cues, beyond the spatial context. These encompass but are not limited to: imaging geometry, imaging mode, properties of the Fourier spectrum of the images or the behavior of the polarimetric signatures. In this paper, we propose a change of paradigm for explainability in data science for the case of Synthetic Aperture Radar (SAR) data to ground the explainable AI for SAR. It aims to use explainable data transformations based on well-established models to generate inputs for AI methods, to provide knowledgeable feedback for training process, and to learn or improve high-complexity unknown or un-formalized models from the data. At first, we introduce a representation of the SAR system with physical layers: i) instrument and platform, ii) imaging formation, iii) scattering signatures and objects, that can be integrated with an AI model for hybrid modeling. Successively, some illustrative examples are presented to demonstrate how to achieve hybrid modeling for SAR image understanding. The perspective of trustworthy model and supplementary explanations are discussed later. Finally, we draw the conclusion and we deem the proposed concept has applicability to the entire class of coherent imaging sensors and other computational imaging systems.
\end{abstract}


%
\IEEEpeerreviewmaketitle

\section{Motivation and Significance}

The Earth is facing unprecedented climatic, geomorphologic, environmental or anthropogenic changes, which require global scale, long term observation with Earth Observation (EO) sensors. SAR sensors, due to their observation capability during day and night and independence on atmospheric effects, are the only EO technology to insure global and continuous observations. Meanwhile, the SAR observations of Sentinel-1 satellites in the frame of the European Copernicus program, are worldwide freely and openly accessible. This is immensely enlarging the SAR Data Science and applications, covering a multitude of areas as: urbanization, agriculture, forestry, geology, tectonics, oceanography, polar surveys, or biomass estimation, only to enumerate a few. Copernicus Open Access Hub provides more than 457.59 PB data of satellites covering the Earth  for more than 570,000 users all around the world. \footnote{https://scihub.copernicus.eu/reportsandstats/}

\begin{figure}[!t]
\centering
\includegraphics[width=8.5cm]{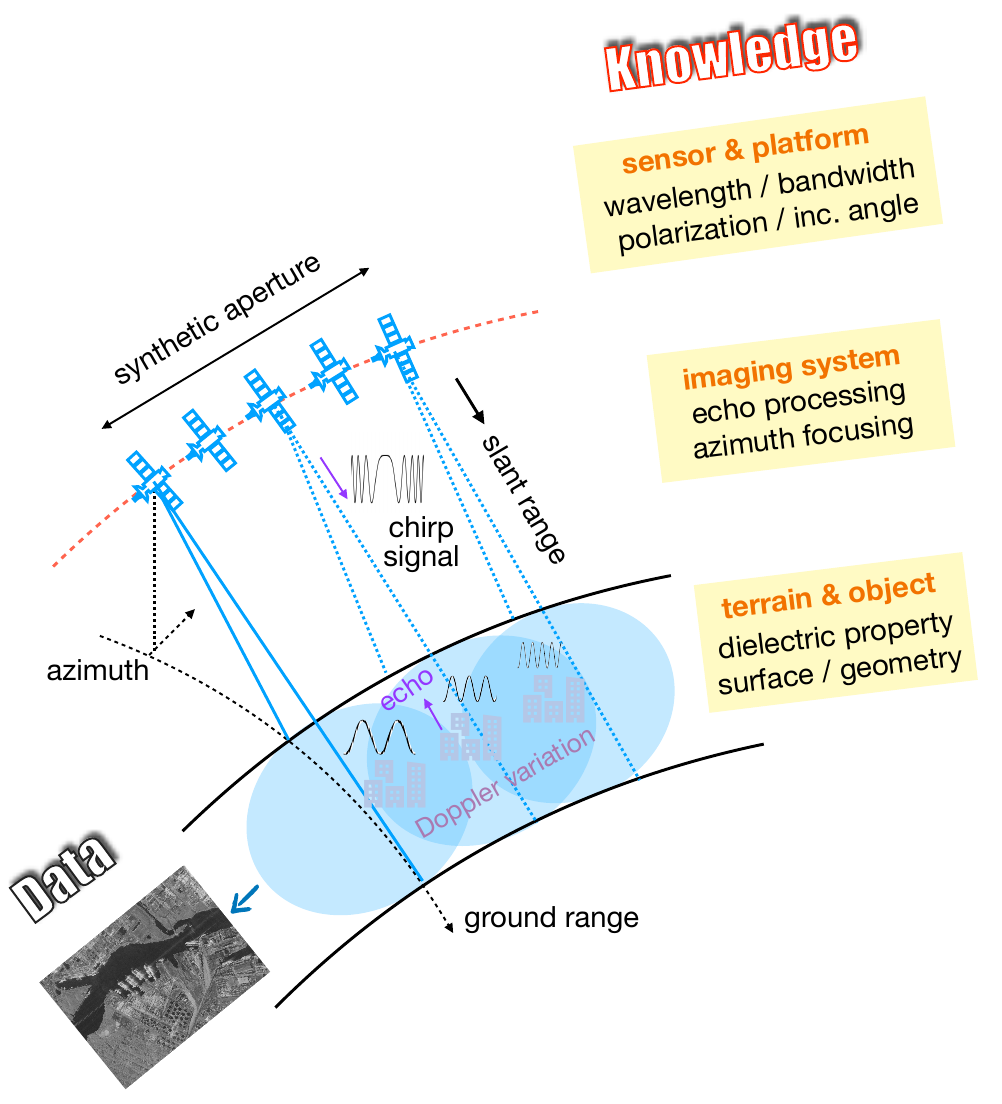}
\caption{A simple illustration of how SAR images the world (Stripmap Mode). SAR society is facing the big data challenge but with limited ground truth. In the meanwhile, the knowledge of SAR is equally important. This is also the motivation of the physical layers in this paper.}
\label{fig_intro}
\end{figure}

SAR is a pioneer technology in the field of computational sensing and imaging, of which the imaging mechanism is totally different from optical sensors. A radar instrument carried by an airborne or spaceborne platform illuminates the scene by side-looking or forward-looking, which allows to discriminate objects in the range direction. As the platform moving along its track, the SAR sensor is constantly transmitting a sequence of chirp signals and receiving echos reflected from objects on the ground, as depicted in Fig. \ref{fig_intro}. When recording all individual acquisitions with a short physical antenna and mathematically combining them into a synthetic image, a much larger synthesized aperture is formed. This allows high capacity to distinguish objects in azimuth despite a physically small antenna \cite{moreira2013tutorial}. A high resolution “image” can be processed by applying SAR focusing principle, e.g., matching filtering \cite{meta2007signal}. 


\begin{figure*}[!t]
\centering
\includegraphics[width=14cm]{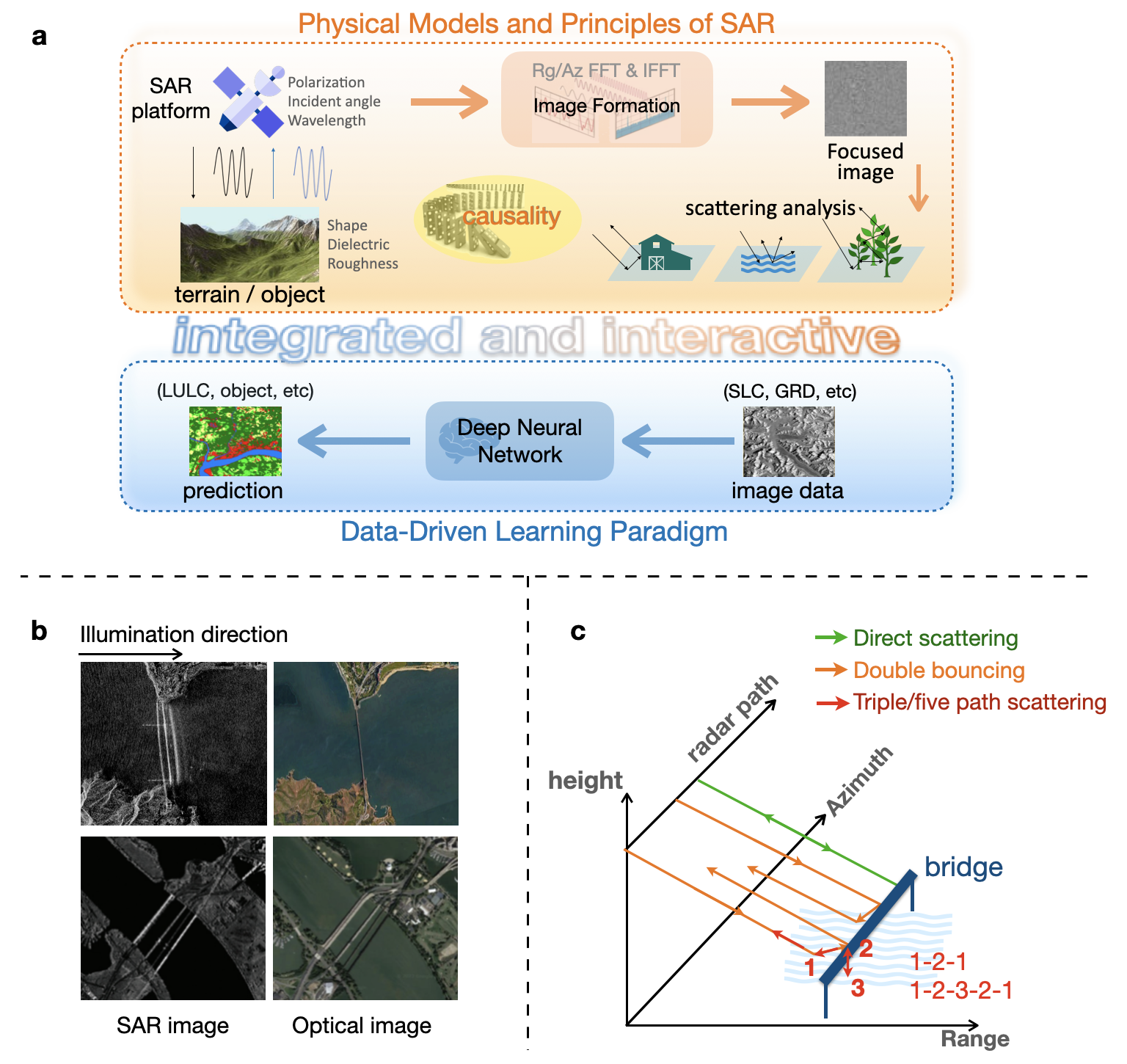}
\caption{\textbf{a} The conventional data-driven paradigm for intelligent SAR image understanding based on deep neural networks and the proposed paradigm shift integrated and interacted with physical knowledge of SAR. \textbf{b} A bridge can be imaged as multiple bright lines, similar as a couple of bridges imaged in the other SAR image, depending on the observation parameters and orientations. This positions the load and outmost difficulty of SAR image understanding. \textbf{c} The multipath scattering formation in the SAR image.}
\label{fig_overview}
\end{figure*}

A deluge of SAR sensors have increased the data availability for various SAR applications. The allure of data-driven learning stems from the ability of automatically extracting abstract features from large data volumes \cite{amrani2021new,amrani2018sar,amrani2017efficient,amrani2017deep}, and therefore, many deep learning studies for SAR applications have been developed in recent years \cite{Zhu2021,rs11131532,HUGHES2020166,Xiang2022}. Current popular paradigm predominantly follows the blue part in Fig. \ref{fig_overview} (a), where SAR image data is all that is required to operate an intelligent network. In addition to data, however, the physical model and principles of SAR sensor should not be neglected. In the upper example of Fig. \ref{fig_overview} (b), A bridge over a placid river that is illuminated perpendicular to its primary orientation appears as many brilliant lines, resembling the lower SAR image in which several bridges are imaged from a different viewing angle. The phenomena can be explicable by multi-path scattering \cite{zhangCharacteristicsMultipathScattering2015,soergelExtractionBridgeFeatures2007}, as illustrated in Fig. \ref{fig_overview} (c). Apart from the direct scattering from the bridge, the double bounce reflection between the bridge and water or vice versa occurs at the corner reflector spanned from the smooth vertical bridge facets facing the sensor and the water surface. In addition, the triple-bounce reflection and maybe some five-path scattering would happen between the horizontal plane of bridge and water surface. Thus, SAR image implies the causality of multi-path scattering phenomena and object characteristics. This positions the load of SAR image understanding, and the outmost challenge of data science, as new and particular paradigm of Artificial Intelligence (AI).

So far, some researches have discussed the paradigm that attempts to bring scientific knowledge and data science models together, applied to a broad range of research themes such as partial differential equation solving \cite{karniadakisPhysicsinformedMachineLearning2021a} and Earth sciences \cite{karpatneTheoryGuidedDataScience2017a,karpatneMachineLearningGeosciences2019,Reichstein2019}. In particular for SAR community, however, this topic has rarely been systematically analyzed and illustrated. Thus, we aim to prospect the hybrid modeling paradigm for intelligent SAR image understanding, where deep learning is integrated and interacted with SAR physical models and principles, to achieve explainability, physics awareness, and trustworthiness.


\begin{figure*}[!htb]
\centering
\includegraphics[width=12.5cm]{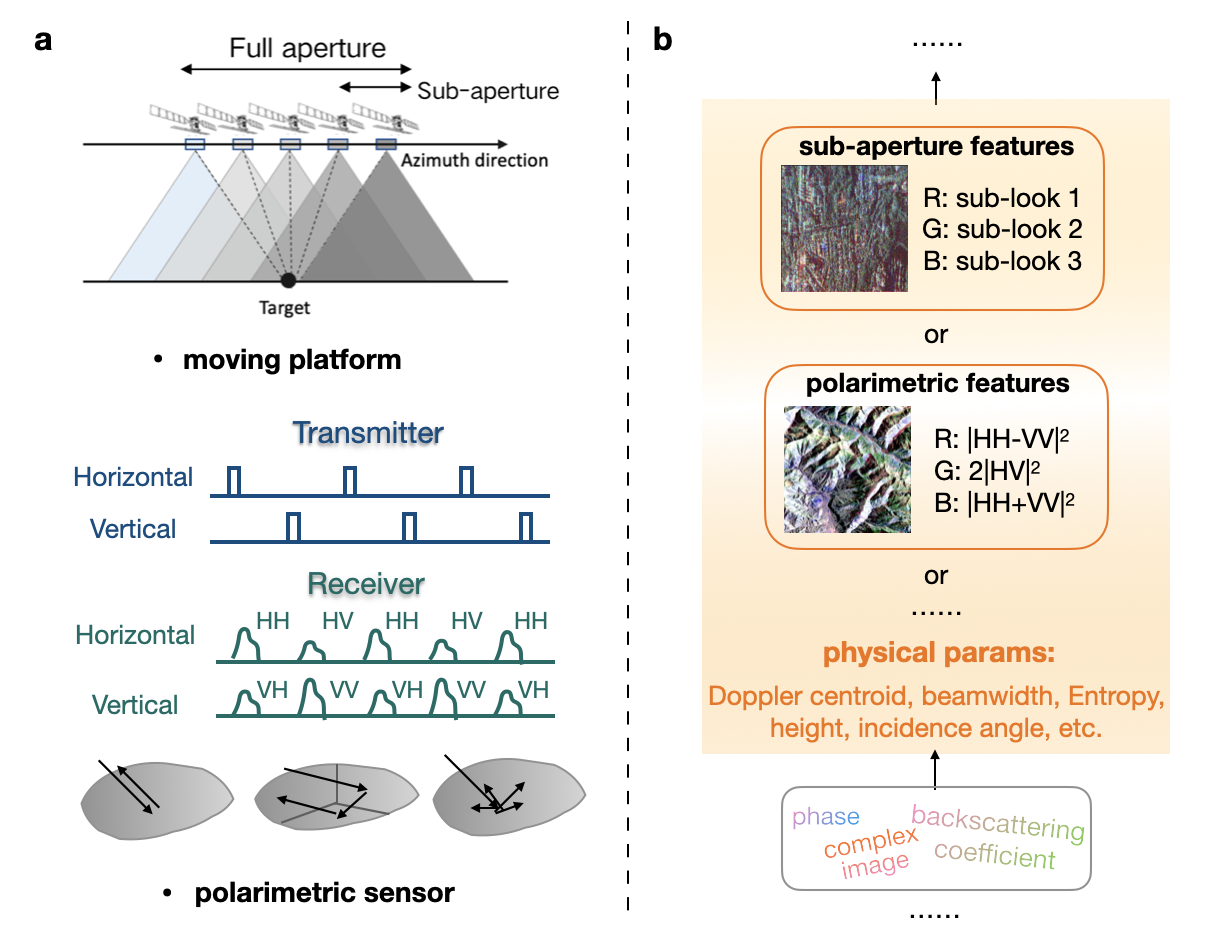}
\caption{Physical layer (i): Sensor and Platform. \textbf{a}: The moving platform creates Doppler variations and synthesizes large virtual aperture; PolSAR transmits and receives diverse polarized wave, and SAR polarimetric characteristics are depicted. \textbf{b}: Based on the physics behind the platform and sensor, the physical layer produces SAR specific representations such as sub-aperture synthesis image and polarimetric feature, with specified physical parameters.}
\label{fig_physicallayer1}
\end{figure*}

Explainable AI is a broad concept. A scientific understanding of explainability is the capacity to clarify the results in the context of domain knowledge. The algorithms still remain a black box. A different approach is the algorithmic explainability. This is constructed such that the results of the used model can be described algorithmically. To obtain a higher degree of explainability, we aim at the synergy of the paradigms: \textit{algorithmic and scientific explainability}. Algorithmic explainability lies in the guarantee of transparency to understand how the machine learning algorithm works by participation of SAR physical models and principles. Scientific explainability ensures the physical consistency of AI output, as well as learning of trustworthy results with physical meaning.

To ground this, we first lay out a representation of SAR physical layer in the context of SAR domain knowledge, as presented in Section \ref{sec:phylayer}. Further, we describe how to integrate and interact them with popular neural networks to build a hybrid and translucent model for SAR applications using illustrative examples, demonstrated in Section \ref{sec:hybrid}. The perspective of trustworthy models and supplementary explanation for SAR community are discussed in Section \ref{sec:trust} and \ref{sec:explanation}. The conclusion and perspectives are finally given in Section \ref{sec:conclusion}.

\section{SAR Physical Layers}
\label{sec:phylayer}

Other than the neural network layers equipped with a number of learnable parameters, SAR physical layers are ones embedded with physical knowledge of SAR, well-established, interpretable, and supported by domain theories. The concept of "physical layer" apart from "neural network layer" arose in literature \cite{Reichstein2019} to make the model more physically realistic. As motivated in Fig. \ref{fig_intro}, three SAR physical layers are highlighted specific for SAR applications in this paper, i.e., (i) sensor and platform: referring to antenna characteristics and moving satellite/aircraft, (ii) imaging system: figuring image formation with focusing process and (iii) scattering signature: reflecting the physical properties of terrain and objects. 


\subsection{Sensor and Platform}

Fig. \ref{fig_physicallayer1} demonstrates the physical layer of sensor and platform that indicates the physics behind the SAR acquisition principle, such as aperture synthesizing with moving platform and various characteristics of antenna.

Existing spaceborne EO SAR missions work in a monostatic or quasi-monostatic configuration. The simplest illumination mode of a SAR system is the stripmap mode in which the antenna pointing direction is constant throughout the acquisition, as shown in Fig. \ref{fig_physicallayer1} \textbf{a}. The moving platform leads to a sliding Doppler spectrum that impacts the complex SAR image. Knowing the behaviour of the Doppler centroid to create sub-looks is essential for exploiting look angle diversity of the input data, especially for very high-resolution SAR images.

\begin{figure*}[!t]
\centering
\includegraphics[width=16cm]{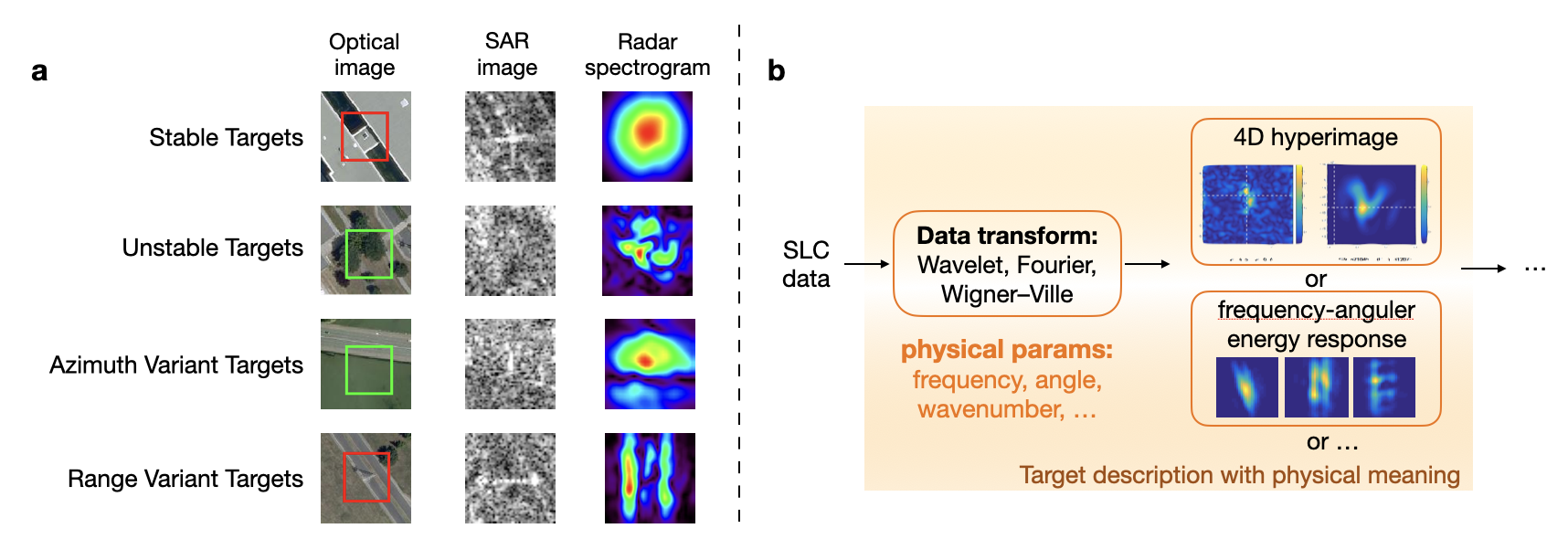}
\caption{Physical layer (ii): Image Formation. \textbf{a.} Targets are characterized by sliding bandpass filtering in the Fourier domain. \textbf{b.} On the basis of image formation principle and target scattering model, the physical layer generates the rich target description with physical meaning.}
\label{fig_physicallayer2}
\end{figure*}

It is well-known that in high-resolution SAR image where the signals are performed over a broad bandwidth and wide angular aperture, the targets are no longer isotropic and non-dispersive. Instead, it is more plausible to infer that the target's backscattering is dependent on illumination angle and frequencies \cite{triaDiscriminatingRealObjects2007}. The sub-aperture processing can be applied to analyze the target scattering variations. Fig. \ref{fig_physicallayer1} \textbf{b} gives an example of a synthesized pseudo color SAR image via sub-aperture processing. The complex-valued SAR image is first transformed to the azimuth spectral domain by a one-dimensional Fourier transform. Then, the full Doppler spectrum is equally split into three intervals, named sub-apertures or sub-looks, each containing 1/3 range of azimuth angles. Finally, the three sub-apertures are transformed back to time-domain using an inverse Fourier transform, coded as the R, G, and B channels, respectively. Red, Green, and Blue targets respond mainly on the first, second, and third sub-looks, respectively, whilst gray targets indicate that they respond equivalently in different sub-looks. The pseudo-colored image well demonstrates the particular behavior of some targets. Given the precise knowledge of the parameters related to Doppler variations (e.g., orbit, azimuth steering rate, radiation pattern, incidence angle), the physical layer can generate sub-look data deterministically and there is no need to design a neural network that should learn to create sub-looks from various types of SAR training data. 


Sensor characteristics, such as polarization, interferometry and tomography, construct physical layer as well. Fig. \ref{fig_physicallayer1} \textbf{b} presents a Pauli pseudo RGB image, where R, G, and B channels are formed with $|HH-VV|^2$, $2|HV|^2$, and $|HH+VV|^2$, respectively, indicating the polarimetric relation. Several physical layers can be stacked to represent rich physics of SAR sensor and platform. Early in literature \cite{ferro-famil_scene_2003}, the diversity in the polarimetric features with the azimuthal look angle was exploited. Thus, the moving platform and polarimetric sensor are both characterized. Similarly, the stacked physical layers can represent polarimetric and interferometric properties of PolInSAR data, or any other combinations.


\subsection{Imaging System}

The second physical layer we suppose delineates the physics behind SAR image formation with an imaging system. The selected exemplars are illustrated in Fig. \ref{fig_physicallayer2}.

A pulse-based radar or a frequency modulated continuous wave (FMCW) radar is usually used in a SAR system, where a range profile is obtained for each transmitted/received waveform, either by range compression in the case of a pulse-based radar or by applying a Fourier transform to the beat signal in the case of an FMCW radar \cite{anghel2014short}. By a coherent processing of the range profiles, the azimuth focusing process outputs a SAR image representing a two-dimensional complex reflectivity map of the illuminated area. SAR processing, taking a simple point target as example, aims to collect the dispersed signal energy in range and azimuth into a single pixel. Many traditional imaging algorithms are in terms of a Fourier synthesis framework \cite{oliver1989synthetic}, as such, Fourier transform provides a specific physical meaning for SAR image. This kind of physical layer assists AI model to better depict the target scattering beyond the "image" domain. 



Fig. \ref{fig_physicallayer2} (a) first shows a simple time-frequency analysis of target with short-time Fourier transform \cite{spigai2008time,singh2011sar}, characterizing the backscattering intensity variations in 2-D range and azimuth frequency domain. Four kinds of backscattering behaviors observed in SAR were defined in literature \cite{Spigai2011}, related to different objects shown in Fig. \ref{fig_physicallayer2}. In the high-resolution case (wide bandwidth chirp signal and broad angular aperture), the complex amplitude of a target is frequency and aspect dependent \cite{triaDiscriminatingRealObjects2007}. Thus, the image formation can be extended to four dimension (called hyperimage) with wavelet transform, providing a concise physically relevant description of target scattering. This frequency and angular energy response pattern is proved useful for discriminating different scatterers, offering valuable prior information to AI model, depicted in Fig. \ref{fig_physicallayer2} \textbf{b}.



\begin{figure*}[!t]
\centering
\includegraphics[width=16cm]{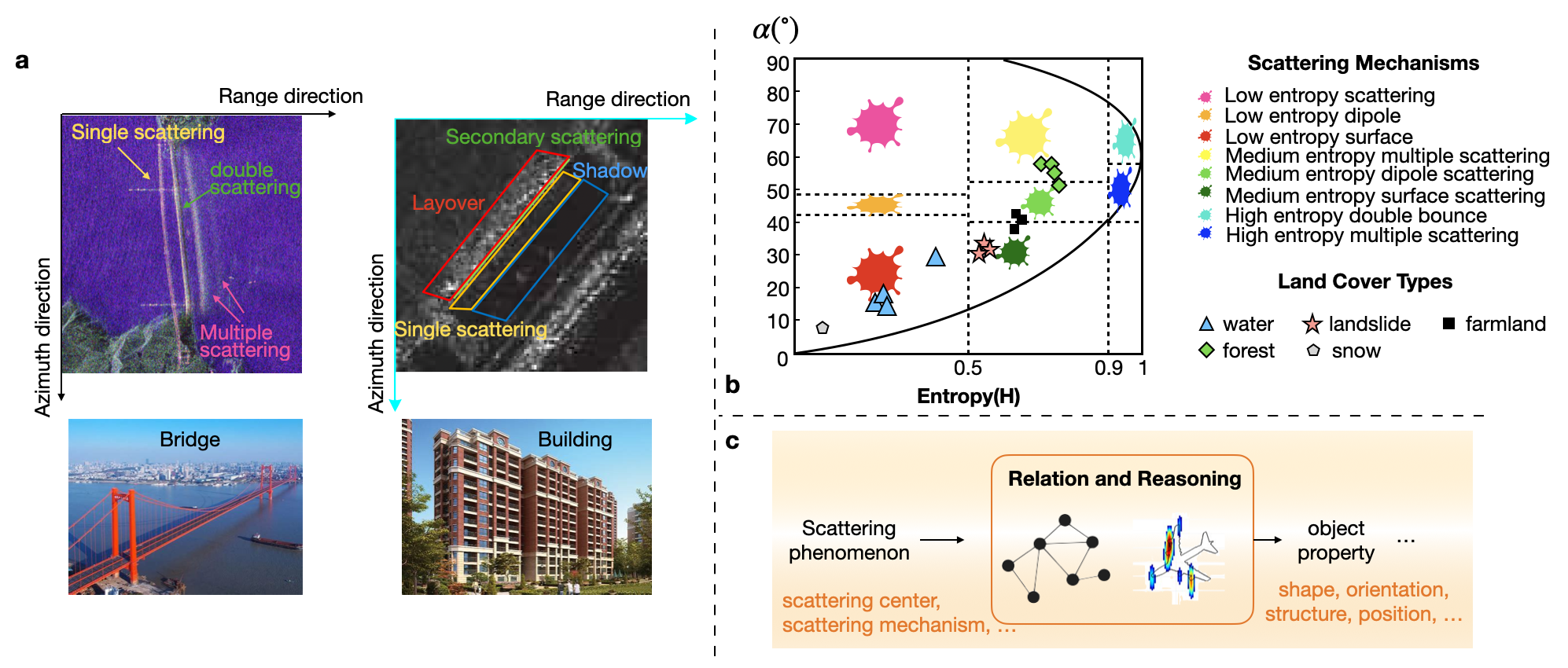}
\caption{Physical layer (iii): Scattering Signatures of Objects. \textbf{a}. The Golden Gate Bridge revealing multipath scattering characteristics in a Gaofen-3 quad-pol SAR image, and a typical single building representing different scattering regions in a high-resolution (1m) SAR image \cite{Chen2021}. \textbf{b}. The scattering mechanisms indicated by the H/$\alpha$ plane for full-polarized SAR data point out the land-use and land-cover classes \cite{Yonezawa2012}. \textbf{c}. The physical layer describes the relationship and reasoning between the scattering characteristics seen in the SAR image and the object's features, such as its shape, structure, or even semantics.}
\label{fig_physicallayer3}
\end{figure*}

\subsection{Scattering Signatures of Objects}

Thirdly, we introduce the physical layer regarding the scattering signatures of objects, in which the causality of target characteristics and scattering behaviors is involved.

For optical images, what you see is what you receive, that is, the objects depicted on the optical image are in accord with human cognition. Targets in SAR images are reflected by scattering characteristics, yet they include a wealth of physical information that the human eye cannot immediately identify. Fig. \ref{fig_physicallayer3} \textbf{a} shows example of two typical SAR targets of bridge and building. The scattering phenomenon that shows several parallel lines over the river can be interpreted as single, double, and multiple scattering of the bridge based on the domain knowledge. The building, with scattering signatures of layover, shadow, single and secondary scattering in the high-resolution SAR image, can also be reflected as only layover and shadow \cite{Chen2021}, depending on the building orientation and shape. Similar research by Ferro et al.  \cite{Ferro2011} investigated the relationship between double bounce and the orientation of buildings in VHR SAR images. Fig. \ref{fig_physicallayer3} \textbf{b} demonstrates the relations between the scattering mechanism of H/$\alpha$ plane and the semantics of land-cover and land-use classes \cite{Yonezawa2012}. Likewise, one can deduce the scattering center position and the specific shape of distributed target from a SAR image by applying some scattering models \cite{Potter1997}.

The conventional data-driven convolutional neural network can capture the image contents as we "see" in the SAR image, whereas it is not equipped with the ability to "interpret" the scattering phenomenon as we discussed before. This indicates the knowledge gap between SAR scattering signatures and human vision cognition. The physical layer delivering semantic understanding behind the SAR scattering signature permits a more thorough interpretation of the SAR image. As shown in Fig. \ref{fig_physicallayer3} \textbf{c}, the physical layer defines the association between the scattering characteristics of a SAR image and the object's qualities, such as shape, structure, or semantics. It can be written as an objective function or a regularization term that constrains the training of neural networks. This will improve the intelligence of AI model to master some causality between scattering signatures and the object nature.

\section{Hybrid Modeling with SAR Physical Layers}
\label{sec:hybrid}

The integration and interaction of neural network layers and physical layers construct the hybrid modeling for SAR image interpretation. In view of algorithmic explainability, the explainable physical models and domain knowledge improves the transparency. For scientific explainability, the hybrid modeling ensures the physical meaning of output in physical layers and the prediction can maintain the physical consistency. In this section, we demonstrate several hybrid modeling approaches with SAR physical layer to achieve explainability and physics awareness.

\begin{figure*}[!t]
\centering
\includegraphics[width=13cm]{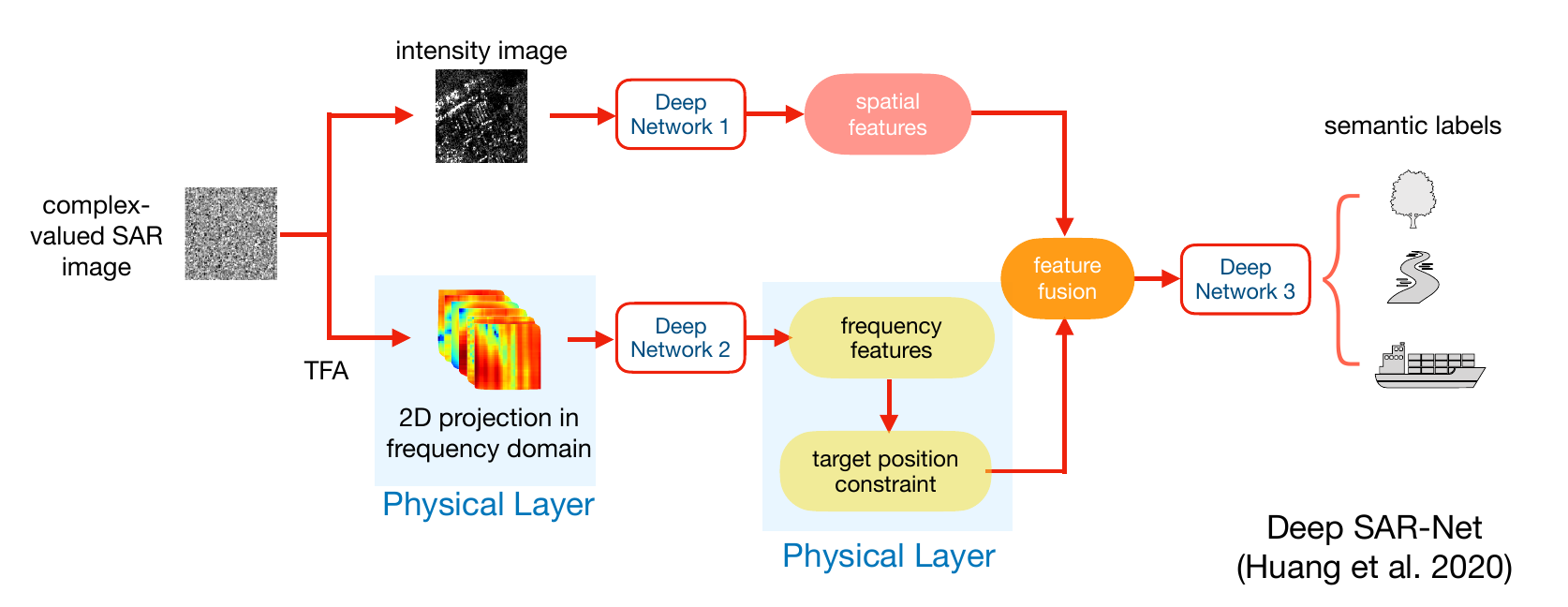}
\caption{Our recent work Deep SAR-Net (DSN) \cite{HUANG2020179} for SAR image classification can be regarded as a typical example of inserting the physical layers into a deep model.}
\label{fig_physicallayer1_eg}
\end{figure*}

\begin{figure*}[!t]
\centering
\includegraphics[width=15cm]{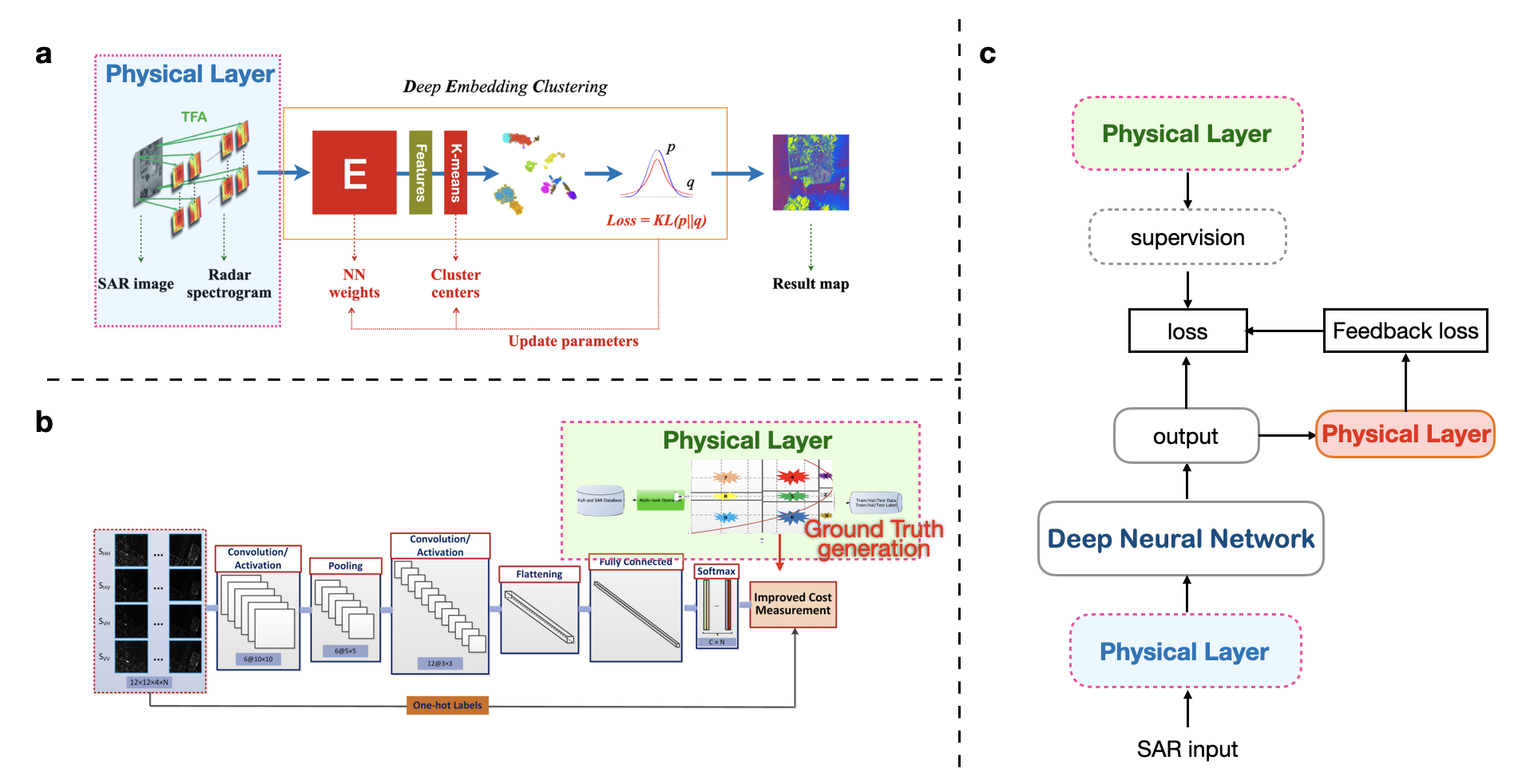}
\caption{\textbf{a}. Unsupervised HDEC-TFA method \cite{Huang2020}. It automatically discovered the radar spectrogram patterns more than the four defined in \cite{Spigai2011} with deep neural networks. \textbf{b}. Learning the polarimetric features from single-polarized SAR image, supervised by Entropy-Alpha-Anisotropy generated from full-pol data \cite{zhao2019contrastive}. \textbf{c}. The physical layers in \textbf{a} and \textbf{b} play the role of input transform (blue) and supervision generation (green) in hybrid modeling. In addition, the physical layer (red) can act as feedback to restrict learning and produce physically consistent outcomes.}
\label{fig_physicallayer2_eg}
\end{figure*}

\subsection{Insert for Substitution}

The introduced physical layer can be inserted in a deep neural network for substitution, extracting explainable and meaningful features, either as the input of a DNN or fused with DNN features in intermediate layers. A common way is to insert a physical layer into the input layer to obtain the polarimetric features for PolSAR image classification, including the elements of coherency matrix, Pauli decomposition features, etc \cite{Zhang2017,vinayaraj_transfer_2020}. Similarly, the sub-aperture images are generated as the input for target detection \cite{lei_feature_2021}. The other usage of physical layer is for feature fusion, where the features obtained by well-established physical model and deep neural networks are combined \cite{Zhang2020,feng_sar_2021}.

Our recent work, a deep learning framework named Deep SAR-Net (DSN) \cite{HUANG2020179}, addressed both aspects that inserts the physical layer into the input and the intermediate position of deep model. As shown in Fig. \ref{fig_physicallayer1_eg}, DSN was proposed for classifying SAR images with complex values. Instead of the entire data-driven method, i.e. the complex-valued convolutional neural networks (CV-CNN), the designed DSN encompasses three shallow neural network modules and two physical layers. The first physical layer generates the high-dimensional radar spectrogram based on time-frequency analysis. The second one handles the features of the 2-D projection along the frequency axises \cite{singh2011sar} to maintain the location constraint, making it possible to be fused with spatial features from intensity image. DSN outperformed CV-CNN especially with limited labeled training data, and had a remarkable performance in discriminating the man-made target scenes compared with the traditional CNN. It demonstrates the Fourier process on single-look complex SAR image embedded the knowledge like synthesizing the antenna well characterizes the physical property of SAR target, and the usages of physical layer cut down unnecessary parameters in neural network layers to improve the model performance with limited ground truth.



\subsection{Compensation for Imperfect Knowledge with Feedback}


In condition of unknown/inconclusive physical models or incomplete knowledge, it is difficult to extract perfect physical parameters or physical scattering characteristics of SAR via model-based methods. For instance, obtaining the polarimetric features from dual-pol, or even single-polarized SAR image. Thus, the physical layer interacted with deep neural network take effect.

\subsubsection{Target Character Identification}

\hspace*{\fill}

Some researches have analyzed the energy response pattern in frequency dimensions of target varied in SAR image, and discussed the nonstationary targets \cite{ferro-famil_scene_2003,ovarlezAnalysisSARImages2003}. Spigai et al. \cite{Spigai2011} pointed out four canonical targets with a rough definition shown in Fig. \ref{fig_physicallayer2} a. However, it remains unknown for many complicated scene and objects. Fig. \ref{fig_physicallayer2_eg} show our related work of using physical layer and deep neural network for compensation of imperfect knowledge. The first is the unsupervised hierarchical deep embedding clustering based on time-frequency analysis (HDEC-TFA) \cite{Huang2020}, which was proposed to automatically characterize the radar spectrogram (or the sub-band scattering pattern defined in \cite{Huang2020}) basically in urban area, discovering the various scattering pattern more than the four specific classes defined in \cite{Spigai2011}. It offered a new perspective to describe the physical properties of single-polarized SAR. Furthermore, we used two stacked physical layers to obtain the polarimetric and time-frequency patterns and analyzed with deep neural network in reference \cite{Huang2020IGARSS}.

\begin{figure*}[!t]
\centering
\includegraphics[width=13cm]{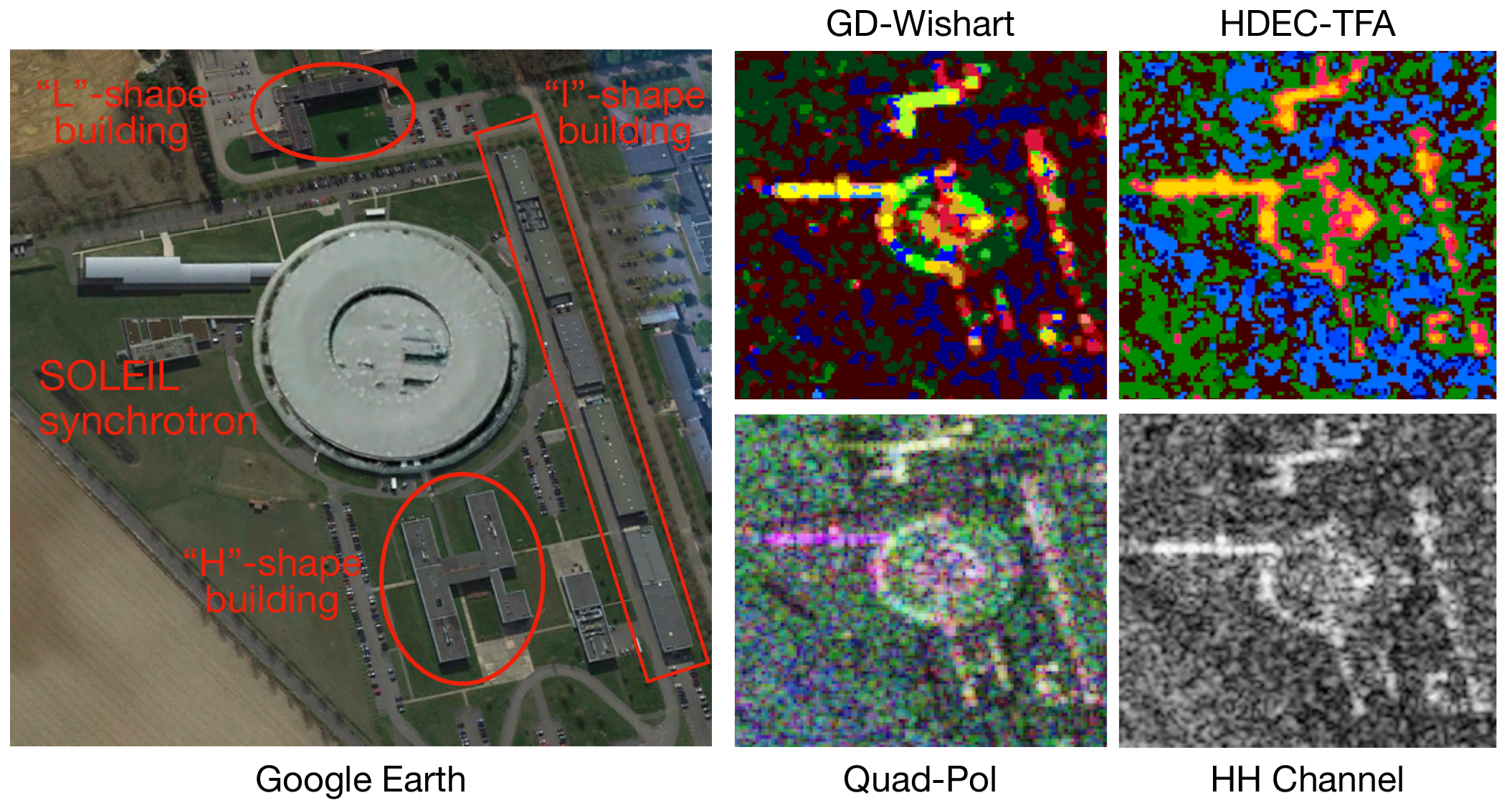}
\caption{The SOLEIL synchrotron in France and the surrounding buildings with different shapes are depicted in the Gaofen-3 SAR image. Both the GD-Wishart \cite{Ratha2018} result on Quad-Polarization SAR data and the HDEC-TFA \cite{Huang2020} result on HH channel single-polarized SAR can capture the special scattering characteristics of the objects.}
\label{fig_HDECTFAeg}
\end{figure*}

Fig. \ref{fig_HDECTFAeg} demonstrates the result compared with the polarimetric physical model. The SOLEIL synchrotron in France, shown as the round building in the Google Earth remote sensing image, is surrounded by three different shapes of buildings. The HDEC-TFA method can capture the special characteristics of the architectures even in single HH channel SAR image, as much as the physical model based method GD-Wishart \cite{Ratha2018} on quad-pol SAR. Some other man-made targets examples characterized by time-frequency model with neural networks are given in \cite{HuangEUSAR2021}. Our experiments in \cite{Huang2020} demonstrated the trained model varies with different imaging conditions since the sub-band scattering pattern is influenced by several imaging parameters, which should be taken into consideration when transferring the AI model to other situations.

\subsubsection{Polarimetric Parameter Extraction}

\hspace*{\fill}

By transmitting and receiving waves that are both horizontally and vertically polarized, the full-pol SAR image captures abundant physical characteristics of the imaged objects that can lead to various physical parameters. In contrast, single-pol and dual-pol SAR data are less informative for physical feature extraction. If only one polarization channel is obtained, one cannot derive the other polarization channels in principle. Once the objects are known, i.e., once the characteristics of targets such as geometry, surface roughness, etc, are identified, deep learning can be employed to transfer the knowledge learned from physical models to reconstruct the physical parameters of objects. As shown in Fig. \ref{fig_physicallayer2_eg} b, Zhao et al. \cite{zhao2019contrastive} proposed a complex-CNN model to learn physical parameters (entropy $H$ and $\alpha$ angle) with transfer learning from single-pol and dual-pol SAR data, supervised by features obtained with a physical layer. Some similar studies include but not limit to \cite{song_radar_2018,qu_study_2021}. Song et al. \cite{song_radar_2018} addressed "radar image colorization" issue to reconstruct the polarimetric covariance matrix with a designed deep neural network, where the supervision was also generated with a physical layer.


When training a data-driven deep neural network, some physical consistencies may not be guaranteed. The authors pointed that the reconstructed covariance matrix may not be positive semi-definite \cite{song_radar_2018}, and they proposed an algorithm to correct it. In this case, the additional physical layer embedded prior constraint acts as post-processing to revise the physically inconsistent result of DNNs. Furthermore, this type of physical layer is suggested to provide feedbacks during training, as demonstrated in Fig. \ref{fig_physicallayer2_eg} c, the red part. The feedback of physical layer aims to prevent the model from learning the physical inconsistency. 

\begin{figure}[!htbp]
\centering
\includegraphics[width=8.5cm]{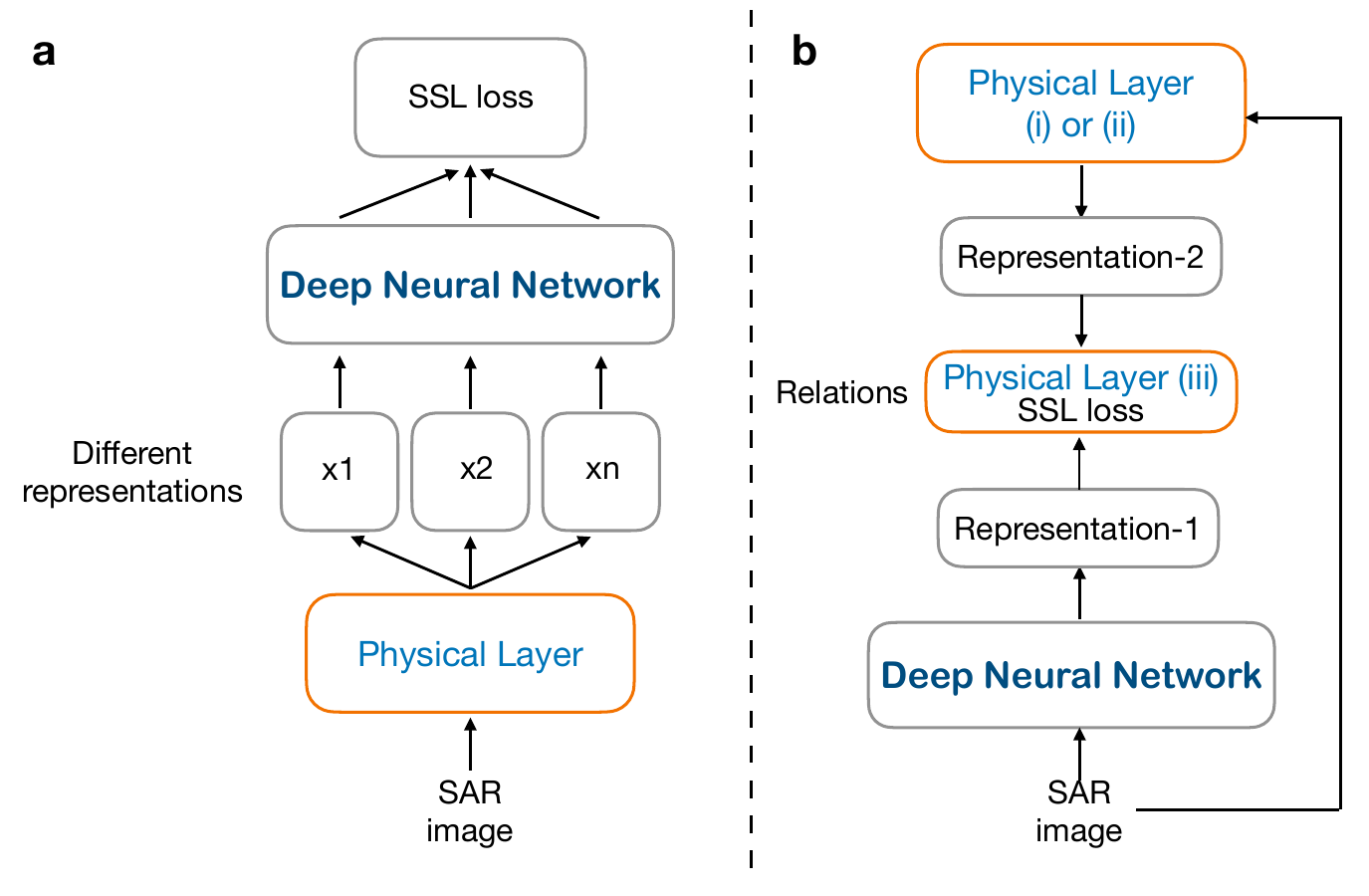}
\caption{The SAR physical layer can be integrated in a self-supervised learning framework to guide the neural network training without ground truth. \textbf{a}. The physical layer generates various modalities of SAR image using well-established physical models, such as sub-aperture images, different polarimetric features, etc. The self-supervised learning can be conducted with contrastive learning paradigm. \textbf{b}. The physical layer produces a physical representation of image, serving as a guided signal that drives the neural network to learn a similar representation.}
\label{fig_ssl}
\end{figure}

\subsubsection{SAR Image Generation/Simulation}

\hspace*{\fill}

\begin{figure*}[htbp]
\centering
\includegraphics[width=12cm]{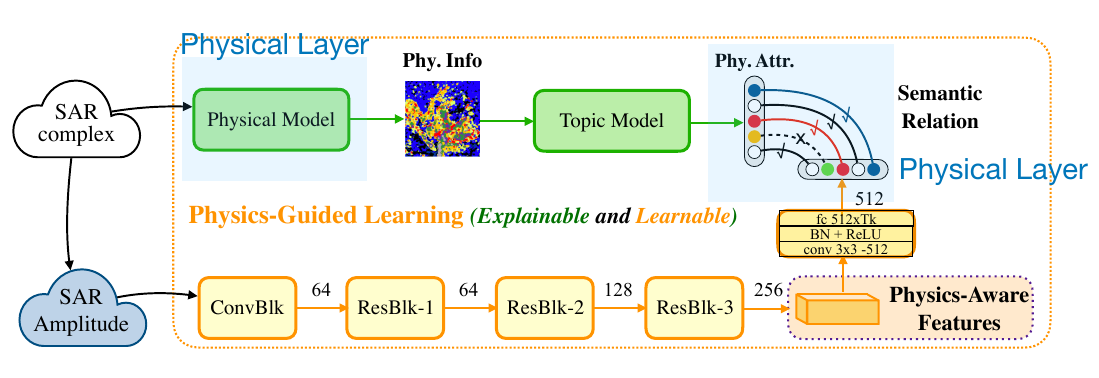}
\caption{A physics guided network was proposed \cite{Huang2021IGARSS,huang2022physically} where a novel deep learning paradigm and loss function were designed to associate the SAR scattering characteristics with image semantics.}
\label{fig_physicallayer3_eg}
\end{figure*}

This paradigm can be popularized to other SAR applications. SAR target image generation (or simulation) based on deep generative model (such as variational auto-encoder \cite{hu_feature_2021} and generative adversarial network \cite{guo2017}) has attracted much attention in recent years. The generated SAR images are expected to be used as data supplements to support target identification. The authenticity and interpretability of current deep SAR image generation is a substantial obstacle that has a significant impact on subsequent tasks \cite{malmgren2017improving}. Many latest studies input important physical parameters into the deep generative model or use them as supervision at the output layer, such as depression angle and target orientation, that facilitated more reliable outcomes \cite{song_sar_2019,oh_peacegan_2021}. We consider them physical layer as shown in Fig. \ref{fig_physicallayer2_eg} c, the green and blue part.

Coupling the physical layer as a feedback in neural network for SAR image generation has yet to be explored. When generative model produces a pseudo SAR image, a physical layer will be applied to verify whether it is consistent with the knowledge base of SAR, e.g. physical parameters derived from a well-established model. If not, the current generative model will revise the pseudo result to minimize the inconsistency. There are some examples to learn from in the field of fluid simulation \cite{chu2021,Xie2018}. As such, the physical layer is used for constructing physical inconsistency as a feedback that explicitly constrain the generative model to fulfill some quantitative conditions, so as to guarantee authenticity. Referred to some latest studies in other fields, physical model as a feedback or constraint in the loop of deep learning is also applied to under water image enhancement \cite{zhouUnderwaterImageEnhancement2022} and seismic impedance inversion \cite{Wang2022physics}.

\subsection{Self Supervised Learning Guidance}


Self supervised learning has been attracted much attention in recent years, since it can help reduce the required amount of labeling. One can pre-train a model on unlabeled data and fine-tune it on a small labeled dataset. It offers great opportunity for SAR community where big data volume is available while the ground truth is usually difficult to obtain. There is a remarkable potential for SAR physical layer to apply for self-supervised learning.

As shown in Fig. \ref{fig_ssl}, two self-supervised learning paradigms are given. The physical layer helps to establish a pretext task for SAR image. In Fig. \ref{fig_ssl} \textbf{a}, different SAR image representations are generated by physical layer, for instance, the sub-aperture images, various polarimetric feature images, etc. As similar to SimCLR \cite{pmlr-v119-chen20j} that conducted the contrastive learning based on data-augmentation, or NPID \cite{Wu_2018_CVPR} that learned the optimal feature via instance-level discrimination, the surrogate task can be built to form a self-supervised learning. An illustrative example is in reference \cite{Ren2021}.

Fig. \ref{fig_ssl} \textbf{b} illustrates a second line of thought, which we refer to as physics guided learning. Firstly, the physical layer is used for generating meaningful physical representations, like scattering mechanisms (physical layer (i) and (ii) can both achieve this). Meanwhile, the neural network extracts hierarchical spatial features from SAR image. The crucial point is how to establish a connection between physical properties and image features. We propose to exploit physical layer (iii) to reveal relationships and thereby design an objective function for self-supervised learning.

Our recent work \cite{Huang2020IGARSS,Huang2021IGARSS,huang2022physically} details the paradigm in Fig. \ref{fig_ssl} \textbf{b}. A physics guided network (PGN) for SAR image feature learning was proposed as shown in Fig. \ref{fig_physicallayer3_eg}. First, a physical layer is deployed at the beginning, where the physical scattering properties are derived. Based on the crucial assumption that SAR image features and the abstract physical scattering mechanisms should share common attributes in semantic level, a surrogate task was established via the other physical layer that defines a loss function. The inspiration is from reference \cite{radu}, which indicated the abstract topic mixture on scattering properties and the high-level image features are with similar semantics. Thus, we built the relation between the image semantics and SAR scattering characteristics. A novel objective function was designed to instruct self-supervised learning guided by physical scattering mechanisms.

The advantages of this kind of learning paradigm lie in two aspects. First, the training process takes all labeled and unlabeled data so that the learned features generalize well in test set. Second, the guidance of physical information leads to physics awareness of features learned by neural networks, i.e., the DNN feature maintains physical consistency. In a word, the prior physical knowledge is embedded in the neural network. The experiments in \cite{huang2022physically} verified this quantitatively and qualitatively. 

\begin{figure*}
     \centering
     \begin{subfigure}[b]{0.43\textwidth}
         \centering
         \includegraphics[width=\textwidth]{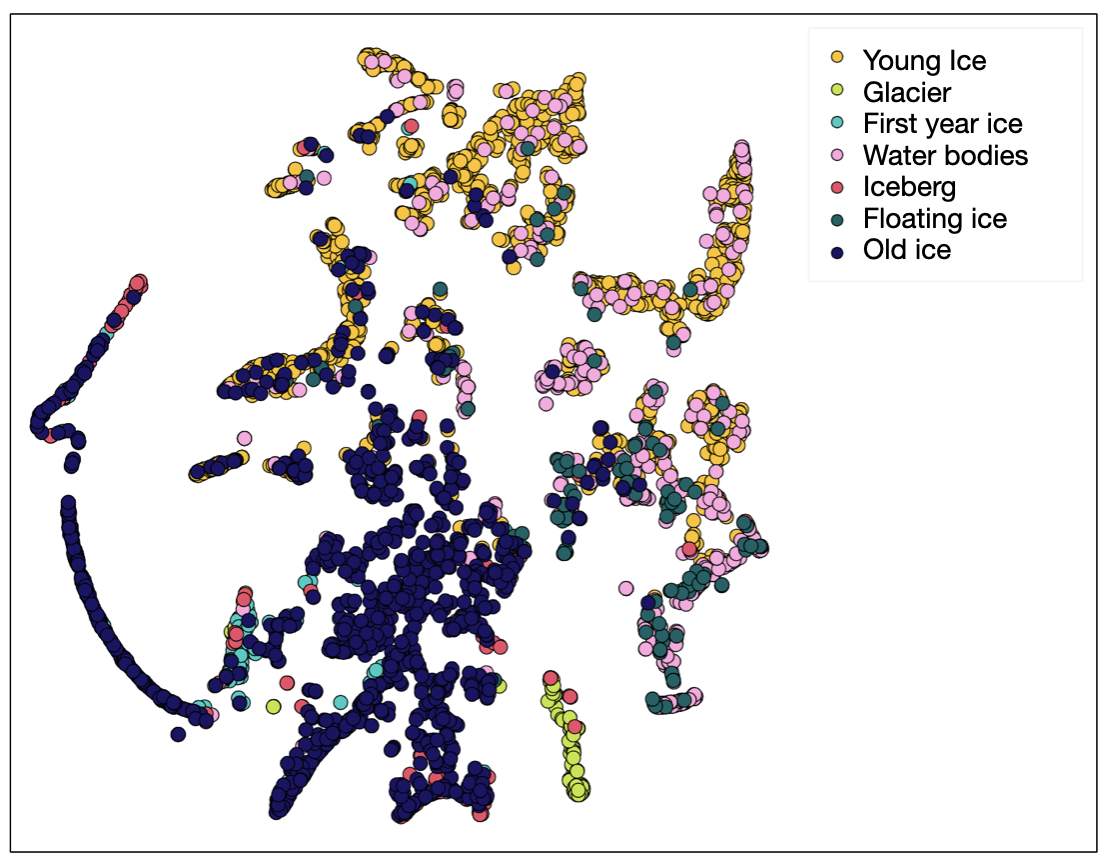}
         \caption{}
     \end{subfigure}
     ~
     \begin{subfigure}[b]{0.43\textwidth}
         \centering
         \includegraphics[width=\textwidth]{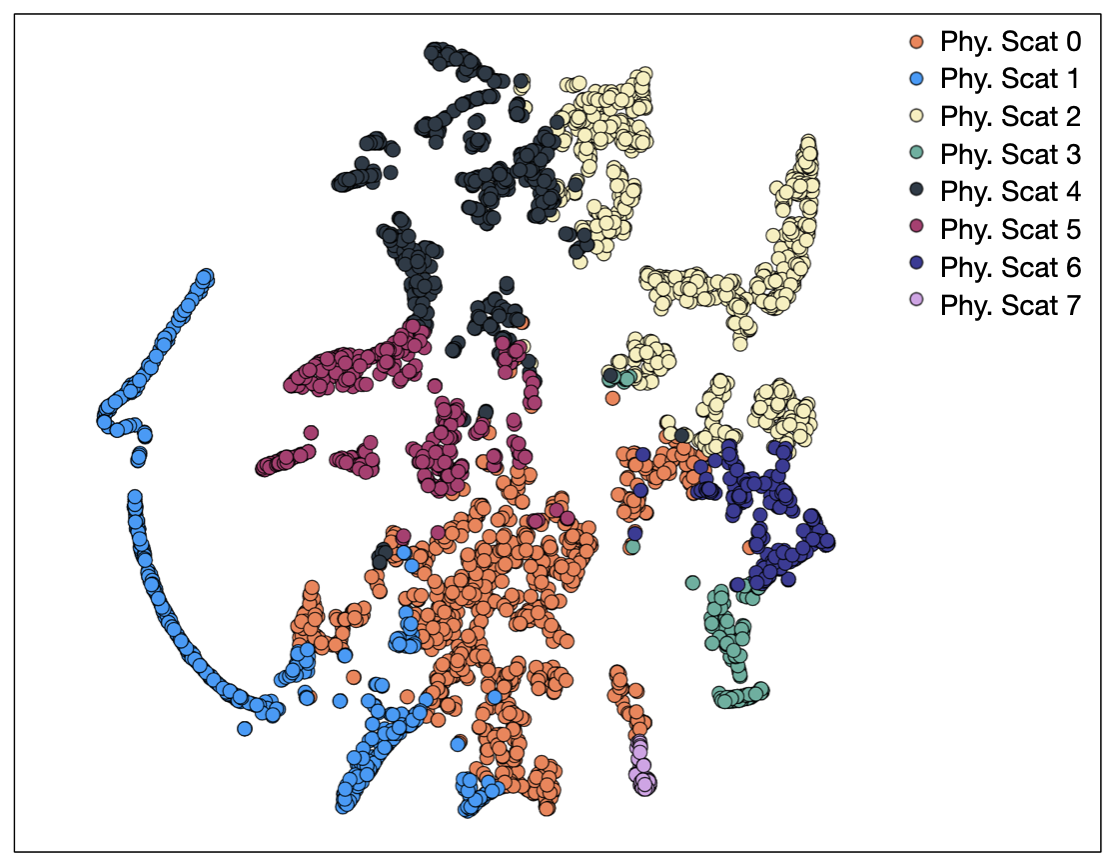}
         \caption{}
     \end{subfigure}
     \caption{Visualization of physics guided signals on test data by t-sne. (a) Different colors represent semantic labels of sea-ice. (b) The physics guided signals are grouped into eight clusters, where each color indicates samples with similar physical scattering properties.}
     \label{fig_featurevis}
\end{figure*}

Additionally, the outputs of the physically interpretable deep model can be further explained, which in turn inspires algorithm improvement. We illustrate with an example of sea-ice classification in polar area \cite{Huang2021IGARSS}. The physics guided learning is driven by physical signals that reflect the scattering properties of SAR image. The guided physical signals are visualized with t-sne in Fig. \ref{fig_featurevis}, where different colors in (a) represent semantic labels of sea-ice and each color in (b) indicates samples with similar physical scattering properties. One characteristic that can be seen is that young ice and water bodies have extremely similar physical representations, which would impede semantic discrimination. It can explain the physics guided learning result in \cite{Huang2021IGARSS} that about 23\% test samples in water bodies class were predicted as young ice. The explanation will motivate us to improve the algorithm by, for instance, relaxing the physical constraints between the two classes.

Similarly, a very recent work \cite{feng2022electromagnetic} was proposed for SAR target recognition inspired by our work\cite{huang2022physically}. The authors proposed a CNN under the guidance of SAR target physical model, attributed scattering center (ASC), to extract the significant target features, that were successively injected into the classification network for more robust and interpretable results.

\section{Trustworthy Modeling}
\label{sec:trust}

\subsection{Why Trustworthy Modeling Needed}

The results obtained by applying AI techniques in SAR processing can be validated using in situ measurements of known targets. For example, a common approach for calibration/validation of SAR data is to employ an electronic target (transponder) that receives a signal, applies a controllable time delay, and transmits the delayed signal towards the receiver of the bistatic/monostatic system. Such a target can be used to validate results related to deformation measurements (e.g., atmospheric corrections) or polarimetric analysis. 

Some real world applications of SAR requires the measurement of reliability and uncertainty. One example is the sea-ice classification in the untraversed polar regions where the ice is always promptly changeable, that would result in the difficulty for annotation and the lack of reference data. In this case, the predictions in unknown polar areas obtained by AI model need to be trusted by humans. Strong robustness and plausible degree of confidence of ML system prediction are equally as important as its accuracy.

\begin{figure*}[!t]
\centering
\includegraphics[width=17cm]{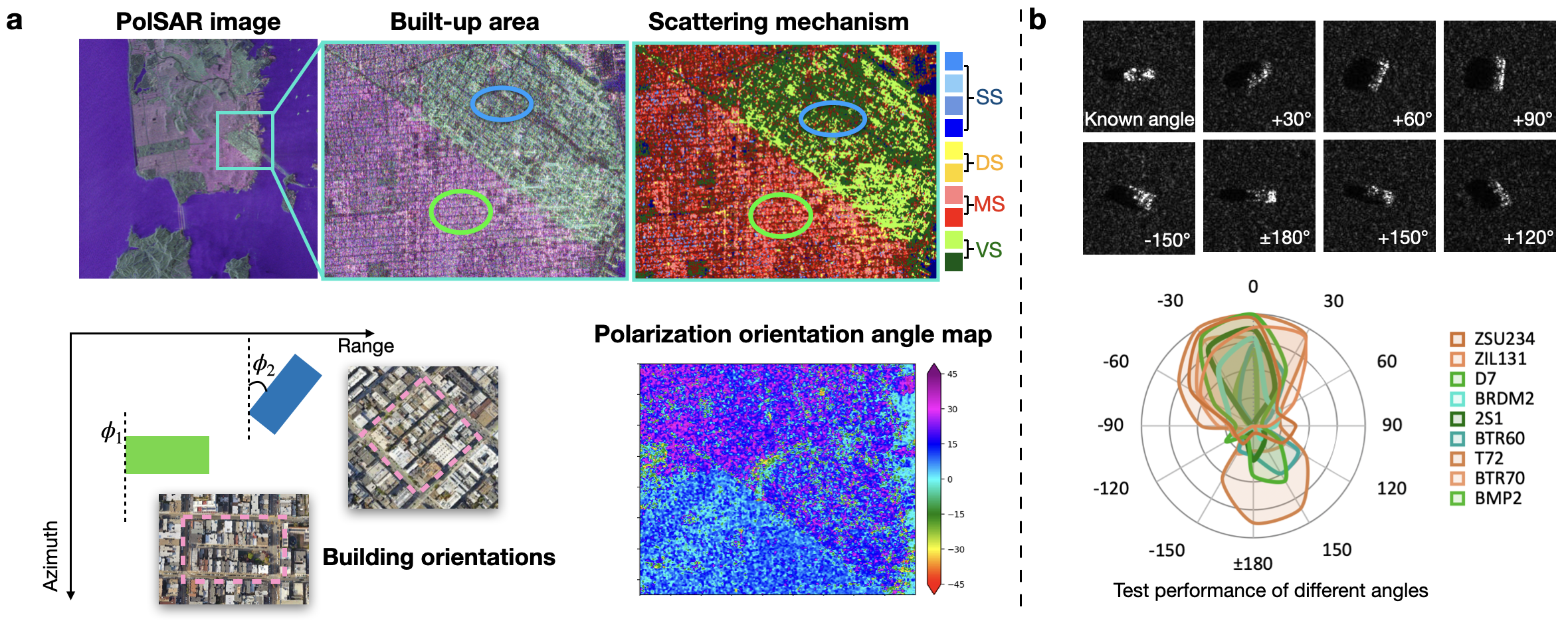}
\caption{A trustworthy model should perceive the SAR scattering variations with a variety of physical parameters and be perturbation-tolerant. \textbf{a} Differently oriented buildings reflect various polarization orientation angles and scattering mechanisms in a PolSAR image. \textbf{b} SAR targets vary violently with orientation angles. When training with a small range of angles, limited robustness of recognition performance is observed as the angle of test data varies.}
\label{fig_trustworthyexample}
\end{figure*}

Fig. \ref{fig_trustworthyexample} \textbf{a} indicates building orientations have a great impact on polarization orientation angles \cite{xiang2016unsupervised} and scattering mechanisms \cite{Ratha2018}. The zoomed-in region mainly contains ortho buildings buildings where $\phi_1$ is close to $0^\circ$ and orientated buildings with a larger $\phi_2$. The polarization orientation angles of ortho buildings are obviously smaller than those of oriented buildings. Ortho built-up areas mainly depict double scattering (DS) and mixed scattering (MS) where the double scattering dominates. The oriented buildings are with volume scattering (VS). Fig. \ref{fig_trustworthyexample} \textbf{b} shows limited robustness of recognition performance as the angle of test data varies when training with a small range of angles. A trustworthy model is expected to perceive SAR scattering variations with a variety of physical parameters and be perturbation-tolerant.

\subsection{Trustworthy Modeling with Uncertainty Quantification}

The development of Bayesian deep learning \cite{wilson2020bayesian} has caught much attention in recent years, where the posterior distribution over parameters are obtained instead of the point estimation. A crucial property of the Bayesian method is its ability to quantify uncertainty, to the benefit of constructing trustworthy model.

In the case of Fig. \ref{fig_trustworthyexample} \textbf{b}, the performance of deep neural networks drops dramatically when testing SAR targets of very different orientation angles with training data. The model is over-confident about some uncertain data that cannot be perceived by frequentist method. Bayesian deep neural network, instead, is able to calibrate the output score and measure the uncertainty of the prediction. Some recent studies applied Bayesian deep learning for SAR sea-ice segmentation \cite{Hartmann2021,Saberi2022,Asadi2021}, as well as target discrimination \cite{Blomerus2021}. The generated uncertainty map can serve as a guideline for the experts in annotation and improve trust between users and the model. Some approximation strategies of Bayesian deep neural network, such as Monte Carlo Dropout \cite{gal2016dropout} and Deep ensembles \cite{lakshminarayananSimpleScalablePredictive2017}, are promising for different SAR applications.

%
%

\begin{figure*}[!t]
\centering
\includegraphics[width=13cm]{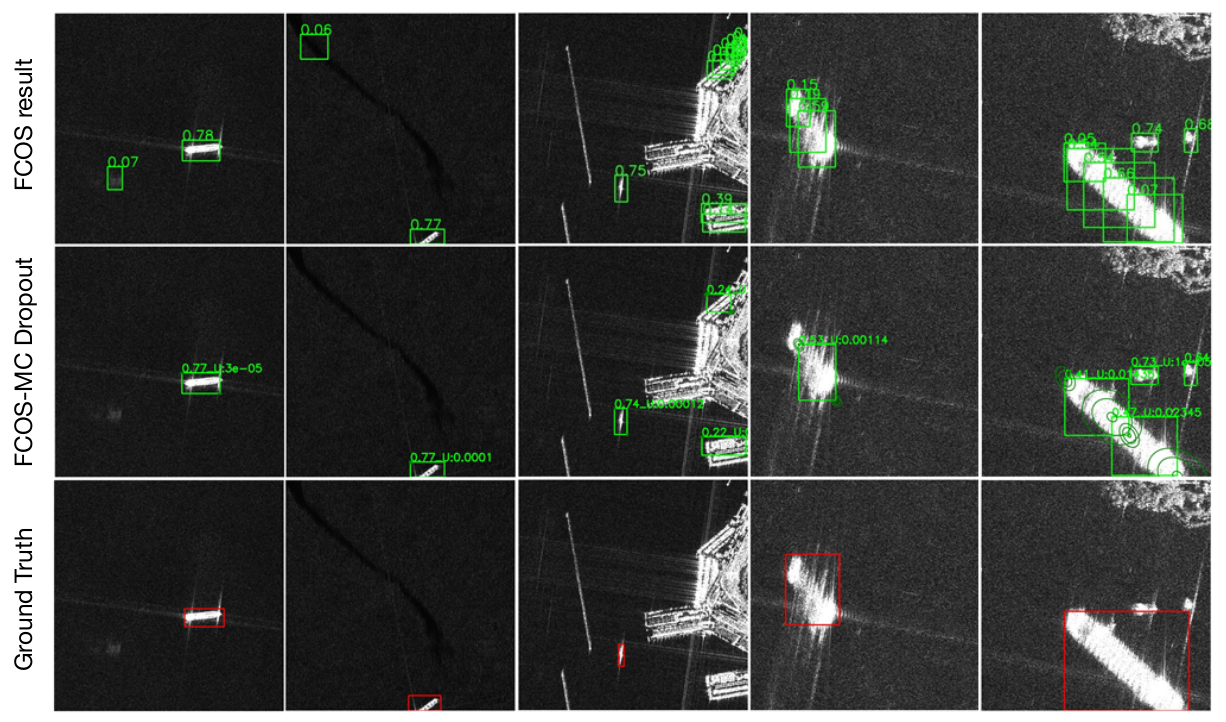}
\caption{The SAR ship detection result on AIR-SARShip-1.0 dataset \cite{R19097}, obtained by the detection deep learning algorithm FCOS \cite{tian2019fcos}. Many false alarms appear in the detection result, due to the limited training data and the interference of complex scattering. The prediction uncertainty is estimated by MC-Dropout \cite{gal2016dropout} and the uncertain results are discard to achieve a better performance.}
\label{fig_uncertaintyresult}
\end{figure*}

We give an example of SAR ship detection for demonstration. The limited labeled training data, and the interference of complex scattering from target itself or the inshore background, would strongly restricted the detection performance. The ship detection result on some selected SAR images from AIR-SARShip-1.0 dataset \cite{R19097}, obtained by FCOS detection algorithm \cite{tian2019fcos}, are shown in the first row of Fig. \ref{fig_uncertaintyresult}. Compared with the ground truth annotation in the third row, the detection result appears many false alarms. It is crucial to estimate the model uncertainty, which is basically brought by inadequate training data, to evaluate the reliability of SAR ship detection model and provide more trustworthy predictions.

When we apply the Monte Carlo (MC) Dropout training strategy to approximate the Bayesian inference \cite{gal2016dropout}, it captures the uncertainty from the existing deep model for SAR ship detection. The results with high uncertainty and very low classification scores are discarded, with only the trustworthy predictions preserved. The results are shown in the second row of Fig. \ref{fig_uncertaintyresult}, where the false alarms are evidently reduced. In the fourth and fifth SAR image, the localization uncertainty of two large ships visualized with circles around the corner of the predicted bounding box is relatively high. Intuitively, we can infer the reason for the weak capability of the trained model in detecting such kind of targets is probably the lack of the large size ships in the training set. The feedback from the uncertainty estimation should further inspire the follow-up studies to improve the algorithm and build trustworthier models.


\section{Supplementary Explanations}
\label{sec:explanation}

Beyond the hybrid and trustworthy modeling, extra explanations and other interpretable models are as well required to assist with developing more transparent AI model for SAR. The explainable artificial intelligence (XAI) techniques, such as gradient based, attention based, and occlusion based explanation methods, are helpful to demonstrate the effectiveness of integrating physical layers to achieve explainability.


The transparent machine learning models, such as linear regression, decision trees, and Bayesian models, are interpretable \cite{BARREDOARRIETA202082}. The algorithm itself provides explanations, for example, Latent Dirichlet Allocation (LDA) builds a three-level hierarchical Bayesian model to describe the underlying relationship among document-topic-word. That is, the document can be explained with a set of topic, where each topic in turn, is represented by a distribution over words. Karmakar et al. \cite{karmakar2020feature} used the LDA model for SAR image data mining to generate the topic compositions and group them into semantic classes, which were fused with domain knowledge obtained by active learning from experts. The transparent model can be also integrated in a deep learning framework to approach the explainability. Huang et al. \cite{Huang2021IGARSS,huang2022physically} applied the LDA model to generate the physical attributes representation as the guided physics signals, rather than directly using the physical scattering characteristic labels to train the physics guided network. That is because the learned physics-aware features are expected to the benefit of semantic label prediction, but the semantic gap actually exists between the physical scattering characteristics and the semantic annotation. Consequently, the LDA model enables the guided signals to gain the abstract semantics and be explained with physical scattering properties.

The other purpose for approaching the explainability lies in the applications of transfer learning. The manual annotation in SAR domain is difficult and the deficiency of labeled data basically restricts the development of data-driven methods. Facing a wide variety of launched SAR platforms with various frequency bands and resolutions, as well as other multi-spectral, hyper-spectral, optical remote sensing sensors, it is of vital importance for elucidating the transferability of ML models among inhomogeneous data. Arrieta et al. \cite{BARREDOARRIETA202082} indicated the transferability is one of the goals toward reaching the explainability. Although many researches have explored different deep transfer learning methods in SAR domain \cite{Huang2017,malmgren2017improving,rostami2019deep}, the inner transfer mechanisms of deep learning model still need explanation of insight. An insufficient understanding of the model may mislead the user toward inappropriate design of algorithm and fatal consequences, i.e. the negative transfer. Based on SAR target recognition, we proposed to analyze the transferability of features in DNN, which contributed to explaining what, where, and how to transfer more effectively for SAR images \cite{huangwhat}. The inspiration also motivates the follow-up studies, including the SAR-specific pretrained model \cite{Huang2021GRSL}, the application in detection task \cite{an2021transitive}, and the interpretability analysis of deep learning model in radar image \cite{LI2021}.


%

\section{Conclusion and Perspectives}
\label{sec:conclusion}

In this paper, we prospect an AI paradigm shift for SAR applications that is explainable, physics aware and trustworthy. To ground this, SAR physical layers embedded with domain knowledge are introduced, which are supposed to be integrated and interacted with neural networks for hybrid modeling. Some illustrative examples are provided to demonstrate the general patterns, showing algorithmic and scientific explainability. In addition, we emphasize the importance and approaches of trustworthy modeling with Bayesian deep learning, as well as illustrating some other techniques such as interpretable machine learning method, explainable techniques, and model transferability, that would assist with developing more transparent AI model for SAR. In fact, this field belonging to interdisciplinary research is still largely undeveloped. To our best knowledge, such approaches have not been formulated in the past years. So far, only some plain attempts have been made. Significant questions and challenges remain, e.g., the feasible representation of SAR physical layer, the optimized form of physical constraint, and hybrid modeling optimization.

Currently there are several smart sensing techniques in the SAR community that can be exploited as pre-processing steps of data fed into DNNs, e.g., multi-aperture focusing in bistatic configurations \cite{rosu2020multiaperture}, monostatic/bistatic tomography, polarimetric decomposition, deformation time series. The outputs of these techniques can expose features that probably cannot be directly extracted by a DNN, especially when using a small training data set.  The newly introduced AI paradigms can apply to the broad class of coherent imaging systems. A few examples can be enumerated: computer tomography, THz imaging, echographs in medicine or industrial applications, sonar or seismic observations in Earth sciences, or radio-telescope data in astrophysics.



%



 \section*{Acknowledgment}

This work was supported by the National Natural Science Foundation of China under Grant 62101459, China Postdoctoral Science Foundation under Grant BX2021248, the Fundamental Research Funds for the Central Universities under Grant G2021KY05104, and a grant of the Romanian Ministry of Education and Research, CNCS - UEFISCDI, project number PN-III-P4-ID-PCE-2020-2120, within PNCDI III. We would like to thank the associate editor and the anonymous reviewers for their great contribution to this article.
\ifCLASSOPTIONcaptionsoff
  \newpage
\fi



%


\bibliographystyle{IEEEtran}
\bibliography{IEEEabrv,SPM}

%



\begin{IEEEbiography}{Mihai Datcu}
(DLR - German Aerospace Center (DLR), Oberpfaffenhofen, Wesling 82234, Germany, and University POLITEHNICA of Bucharest (UPB), mihai.datcu@dlr.de). His research interests include explainable and physics aware AI, smart radar sensors design, and quantum machine learning with applications in Earth Observation (EO). He holds a Professor position in AI and information theory with UPB, and he is Senior Scientist with DLR. He elaborated AI4EO programs and systems for CNES, DLR, EC, ESA, NASA, ROSA. He was the recipient of the Chaire d'excellence internationale Blaise Pascal 2017 for international recognition in the field of Data Science in Earth Observation. He is IEEE Fellow.
\end{IEEEbiography}

\begin{IEEEbiography}
{Zhongling Huang}
(Northwestern Polytechnical University, Xi'an, China. huangzhongling@nwpu.edu.cn). She received the B.Sc. degree in electronic information science and technology from Beijing Normal University, Beijing, China, in 2015, and the Ph.D. degree from the University of Chinese Academy of Sciences (UCAS) and the Aerospace Information Research Institute, Chinese Academy of Sciences, Beijing, China, in 2020. She served as a visiting scholar in German Aerospace Center (DLR) during 2018-2019, funded by UCAS. She is currently working in the BRain and Artificial INtelligence Lab (BRAIN LAB), School of Automation, Northwestern Polytechnical University, Xi’an, China. Her research interests include explainable deep learning for synthetic aperture radar (XAI4SAR), SAR image interpretation, deep learning, and remote sensing data mining.
\end{IEEEbiography}

\begin{IEEEbiography}
{Andrei Anghel}
(University POLITEHNICA of Bucharest (UPB), 313 Splaiul Independentei, Bucharest 060042, Romania, andrei.anghelatmunde.pub.ro). His current research interests include remote sensing, smart radar, microwaves and signal processing. Between 2012 and 2015, he worked as a PhD researcher with Grenoble Image Speech Signal Automatics Laboratory (GIPSA-lab), Grenoble, France. Presently he is Associate Professor in telecommunications at UPB designing bistatic SAR systems for ESA. He is IEEE Senior Member.

\end{IEEEbiography}

\begin{IEEEbiography}
{Juanping Zhao}
(School of Electronic Information and Electrical Engineering, Shanghai Jiao Tong University, Shanghai 200240, China, juanpingzhaoatsjtu.edu.cn). Her research interests include AI for SAR and PolSAR image interpretation, pattern recognition, and machine learning. From 2018 to 2019, she was a visiting Ph.D. student with German Aerospace Center (DLR), Oberpfaffenhofen, Germany.
\end{IEEEbiography}

\begin{IEEEbiography}
{Remus Cacoveanu}
(University POLIEHNICA of Bucharest (UPB), 313 Splaiul Independentei, Bucharest 060042, Romania, rcacoveanu at yahoo.com). His main field of expertise is in the wireless communication systems, antennas, radar sensors, propagation, and microwave circuits. He holds an Associate Professor position in telecommunications with UPB. For more than 10 years he was the technical lead of the Redline Communications’ Romanian branch and between 2011-2015 he was technical consultant for Blinq Networks Canada. One of the designed products received the “Best of WiMAX World EMEA 2008 Industry Choice Award”, Munich Germany May 21st 2008.
\end{IEEEbiography}




\end{document}